\documentclass[twocolumn,showpacs,preprintnumbers,amsmath,amssymb]{revtex4}

\usepackage{graphicx}
\usepackage{dcolumn}
\usepackage{bm}

\begin{document}
\preprint{APS/123-QED}
\title{Universal scaling of forest fire propagation}
\author{Bernard Porterie}
\email{Bernard.Porterie@polytech.univ-mrs.fr}
\author{Ahmed Kaiss}
\author{Jean-Pierre Clerc}
\affiliation{Universit\'{e} de Provence, IUSTI/UMR CNRS 6595, 5 Rue E. Fermi, 13453 Marseille Cedex
13, France\\}
\author{Nouredine Zekri}
\email{nzekri@yahoo.com}
\author{Lotfi Zekri}
\affiliation{ Universit\'{e} des Sciences et de la Technologie d'Oran, D\'{e}partement de Physique,
LEPM, BP 1505 El Mnaouer, Oran, Algeria}
\date{\today}
\begin{abstract}

In this paper we use a variant of the Watts-Strogatz small-world model to predict wildfire behavior
near the critical propagation/nonpropagation threshold. We find that forest fire patterns are
fractal and that critical exponents are universal, which suggests that the
propagation/nonpropagation transition is a second-order transition. Universality tells us that the
characteristic critical behaviour of propagation in real (amorphous) forest landscapes can be
extracted from the simplest network model.

\end{abstract}
\maketitle

Because of heterogeneous conditions of weather, fuel and topography encountered, the spread of
large fires looks fractal, as revealed by satellite maps \cite{caldarelli01}. This suggests
stochastic modeling. Until now this spread has been modelled using regular networks, as well as
cellular automata to include site weights \cite {albinet86}-\cite {barr97}. However there is some
evidence that networks with only local contacts do not mimic real fires very well
\cite{degennes}-\cite{newman02} as they cannot take into account physical effects beyond the
nearest neighbours of a burning site (i.e. an item of vegetation), such as radiation from flames.
Moreover, wind and topography may induce anisotropy, which in turn reduces the effective dimension
of the propagation to less than two. Here we propose a new model based on the social small-world
network ($swn$) which was initially proposed by Watts and Strogatz \cite{watts98} (Fig. 1a).  The
present model is built from a two-dimensional $L-$size lattice which includes both short-range
connections, via the nearest neighbours of a burning site, and long-range connections in its
radiative influence zone (Fig. 1b). The so-called long-range spotting process, whereby burning
firebrands produced when shrubs and trees burn rapidly are lofted by the fire plume and transported
downwind to start new fires in recipient fuel beds, is not considered here as it introduces a
different kind of transition \cite{newman03}. The influence zone of a burning site is characterized
by two parameters, $l_x$ and $l_y$, which are dependent on fire and fuel conditions and expressed
in an arbitrary length unit ($\delta l$) corresponding to the lattice parameter. The vector $\vec
s(s_x,s_y)$ is the position vector of any site in the influence zone with respect to the burning
site position taken as the origin. Isotropic propagation, i.e. with no wind and no slope effects,
corresponds to circular patterns ($l_x=l_y$). A high value of the ratio $AR=l_y/l_x$ corresponds to
a strong anisotropy of the front shape induced by the terrain slope and/or wind effects in the
propagation direction (here the $y-$ direction). Fire patterns then become elliptical. As shown in
Fig. 2, fire patterns exhibit a $swn$ behavior only locally, which explains why the present model
is referred to as a local small-world network ($lswn$). These $lswn$ effects increase as the
proportion of combustible sites, $p$, decreases. The model has shown an excellent agreement with
known experimental data \cite{comb}. In addition to its capacity to mimic the phenomena of fire
spread in nature, the present model produces super-real-time simulations of fire patterns. We then
focus on the very powerful feature of universality of forest fire propagation, as has been done for
the study of complex systems, such as biological systems \cite{gisiger01}.

 Assuming every burning site to be a point-like source of radiation, the
amount of power received by a site located within the influence zone decreases inversely with the
square of the distance from the burning site
 \textbf{
\begin{equation}
\Delta P = \delta P \left(\frac{s_{x}^2}{l_{x}^2}+\frac{s_{y}^2}{l_{y}^2}\right )^{-1}
\end{equation}
} The term $\delta P$ is the minimum amount of power required for a combustible site to start to
degrade. It can be determined from either experiments or deterministic simulations
\cite{porterie07}. The $lswn$ model uses a time-weighting procedure on sites in order to model fire
propagation through a network of vegetation items. It is based on the knowledge of two
characteristic times, namely the time required for a site to achieve complete combustion, $t_{c}$,
and that of thermal degradation before ignition, $t_{TD}$. As an example, the nearest neighbour of
a burning site in the propagation direction receives an amount of power equal to $\delta P (l_y /
\delta l)^2$, whereas a site at the border of the influence zone of this burning site receives an
amount of power $\delta P$ and requires a time of $t_{TD}$ to reach ignition. Fuel properties (e.g.
load distribution, type, or moisture content) do not affect the weighting procedure but the
characteristic times should be modified. The concept of both time-weighting procedure and influence
zone differentiates the $lswn$ model from the usual percolation model, although it has been shown
to behave as a regular network for system sizes much larger than that of the influence zone
\cite{pre}.

 The next point to be considered is the dynamic aspect of the $lswn$ model. At each time step, $\delta t$, taken as
 the time unit, a burning site contributes to the increase in the degradation level
 of the sites connected to it before ignition. Once ignited, these
burning sites each contribute to the thermal degradation and ignition of the sites located in their
own influence zone. Fire cannot propagate if a site ceases to burn before the complete thermal
degradation of its nearest neighbours. The dynamics of propagation in the $y-$direction can be
characterized by the ratio $R$ defined as

\begin{equation}
R=\left(\frac{l_y}{\delta l}\right)^2 \frac{t_c}{t_{TD}}
\end{equation}

There exists therefore a dynamic threshold, below which fire cannot propagate even in a homogeneous
vegetation. For line-ignition conditions, the dynamic threshold value was found \cite{pre} to be
$1/2$ whereas it is unity for point-ignition conditions. For the latter conditions, above the
dynamic threshold, a ratio $t_c/t_{TD}$ smaller than unity means that only a fraction $t_c/t_{TD}$
of the influence zone of the burning site will be ignited when this site finishes burning.

The fire propagation process is initiated here by igniting the site located at the center of the
first line. The statistical averaging process is carried out by generating $N$ samples, $N$ being
ranged from $500$ to $1000$, which ensures that fluctuations are small far from the percolation
threshold.

The geometric propagation (percolation) threshold of the network corresponds to a minimum
concentration of active sites, ($p_{c}$), above which propagation along a path connecting the
opposite sides of the network occurs. In the usual percolation theory \cite{stauf} for infinite
size networks, the correlation length $\xi$, defined as the average cluster size, diverges at $p_c$
as a power law ($\xi \propto (p_{c}-p)^{-\nu}$, $\nu$ being the critical exponent of the
correlation length \cite{stauf}). Near $p_c$, the average burned mass is dominated by that of the
largest cluster which behaves as a power law of $p_c-p$,

\begin{equation}
<m> \propto \left(p_c-p\right)^{-\gamma}
\end{equation}

Below $p_c$, for system sizes $L$ smaller than $\xi$, $<m>$ scales as $(L/ \delta l)^{D_f}$ where
$D_f$ is the fractal dimension of the largest cluster mass and is $p-$dependent. For very large
system sizes, it saturates as $(\xi/ \delta l)^{D_f}$ . The size at which the saturation occurs is
therefore a measure of the correlation length. At $p_c$, the correlation length diverges and the
average burned mass scales as

\begin{equation}
<m(L)> \propto L^{- \gamma/\nu}
\end{equation}

which leads to $D_f(p_c)=\gamma/\nu$.

In the present study we first determined the percolation threshold, the fractal dimension $D_f$ and
the critical exponents as functions of $l_y / \delta l$ for isotropic fire propagations (Fig. 3).
Different techniques can be used to determine the percolation threshold by varying the
concentration of active sites. It corresponds to the peak of the percolation probability
p-derivative \cite{stauf}. It also corresponds to the maximum propagation time as well as the
maximum propagation time fluctuations. These fluctuations are largest at the percolation threshold
since the largest cluster is mainly composed of critical 'red' bonds which stop the propagation if
they are cut \cite{stauf},\cite{comb}. These techniques that are systematically used to determine
the percolation threshold provide very similar values. In the present case, the percolation
threshold decreases as $(l_y / \delta l)^{-1.58}$ (Fig. 3a). The correlation length critical
exponent $\nu$ is in agreement with that of the usual percolation theory \cite{stauf}, $\nu=4/3$
(Fig. 3b), as expected from a re-normalization procedure \cite{loretto} of the isotropic influence
zone. Model results show that for isotropic propagation, long-range radiative effects preserve the
features of the phase transition of the usual percolation theory.

The propagation of real fires is generally anisotropic due to windy and/or hilly terrain
conditions, which enhances overhead flame radiation. As shown in Fig. 4a, the percolation threshold
varies as a power law of the anisotropy ratio, $AR$ (defined in Fig.1), with an exponent of $-2/3$.
Such behaviour has also been predicted for anisotropy crossover in percolation \cite{clerc}. At
$p_c$ the average burned mass scales as $(L/ \delta l)^{D_f}$ with a fractal dimension of the
physical support of fire propagation lower than two (Fig. 4c), whatever the anisotropy ratio
$AR>1$. By analogy with isotropic propagation, the dimension of the propagation support can thus be
estimated within the range of $[1.2-1.5]$. The value of the corresponding correlation length
exponent is approximately $\nu=5/3$ (Fig. 4b), which agrees well with the value extrapolated from
the fit of $\nu(D)$ data obtained for Euclidian system dimensions \cite{stauf} ($2 \geq D \leq 6$)
(Fig. 5). Results summarized in Table 1 confirm the universality of the critical exponents whatever
the network symmetry.

We also examined the behaviour of the rate of spread, $ros$,
defined as the time derivative of the distance covered by the head
fire front at steady state. For homogeneous systems, the $ros$
increases first with the square of the impact length (Eq.1),
whereas during steady propagation it is found to increase as $(l_y
/ \delta l)^{2.6}$ due to the collective contribution of the
neighboring burning sites. For inhomogeneous systems, we propose a
new exponent, $\kappa$, of the scaling law of the $ros$ above the
percolation threshold. To the best of our knowledge, there is no
equivalent exponent in the usual percolation theory. Below $p_c$
fire cannot reach the opposite side and the steady state is never
attained. Above $p_c$ fire spreads at the same rate within large
clusters independently of the system size $L$. For an isotropic
propagation, the critical exponent of the $ros$ seems to be
independent of the impact parameter (Table 1).

A special emphasis is put on the problem of diffusion in the $lswn$ model. The statistical
fluctuations of the distance covered by the front vary with time as $t^{\alpha}$ where the exponent
$\alpha$ is $p-$dependent. This exponent is around $3/4$ for a density $p$ close to $p_c$, which
indicates a super-diffusive propagation \cite{metzler} (it may be remembered that $\alpha=1/2$ for
a diffusive propagation). It becomes unity above the percolation threshold leading to a ballistic
transport, as a result of radiation beyond the nearest neighbours.

Three results emerge from the present study. First, the propagation / non-propagation transition is
a second-order phase transition, the critical exponents being universal. This means that model
results obtained on simple lattices near critical points remain valid for amorphous networks
representative of wildland landscapes. It is found that the range of variation of the $lswn$
percolation threshold is reduced to about $\pm 10\%$ as the network symmetry varies. Second, at the
percolation threshold, fire propagation is super-diffusive. Above $p_c$ it is ballistic. Third, the
dimension of the physical support of anisotropic propagation is fractal, whereas that of isotropic
one is Euclidian. The fractal dimension of the support allowed us to estimate the critical exponent
$\nu$ for fractal dimensions. We are at present examining the role of spotting in the propagation
of wildland fires.
\begin{acknowledgments}
This work was supported by a CNRS grant (ANR PIF/NT05-2-44411).
\end{acknowledgments}
%

\begin{table*}
\caption{\label{tab:wide} critical exponents of fire propagation. $\nu_{th}$ is deduced from
$\kappa =\nu_{th}-1/\nu_{th}$.}
\begin{ruledtabular}
\begin{tabular}{lllcccc}
\hline $l_x/ \delta l$   & $l_y/ \delta l$       &$Lattice$       & $p_c$     & $\gamma$ & $D_f$ &
$\nu$ \\ \hline
 $$ &$$                                  &$Square$        & $0.225$   & $2.26$ ($\pm 0.02$)   & $1.88$ ($\pm 0.05$)   & $1.24$ ($\pm 0.05$)  \\
 $3$ &$3$                                  &$Triangular$    & $0.178$   & $2.19$ ($\pm 0.04$)   & $1.63$ ($\pm 0.04$)   & $1.34$ ($\pm 0.06$)  \\
 $$ &$$                                  &$Amorphous$     & $0.200 $  & $2.09 $ ($\pm 0.04$)  & $1.67 $ ($\pm 0.04$)  & $1.25$ ($\pm 0.06$)  \\
\hline
 $$ &$$                                  &$Square$        & $0.225$   & $2.06$ ($\pm 0.05$)   & $1.26$ ($\pm 0.03$)   & $1.63$ ($\pm 0.08$)  \\
 $2$ &$5$                                  &$Triangular$    & $0.195$   & $2.17$ ($\pm 0.04$)   & $1.34$ ($\pm 0.03$)   & $1.63$ ($\pm 0.07$)  \\
 $$ &$$                                  &$Amorphous$     & $0.210$   & $1.84 $ ($\pm 0.03$)  & $1.12$($\pm 0.02$)    &$1.64$ ($\pm 0.06$)  \\
\hline
 $$ &$$                                  &$Square$        & $0.165$   & $2.53$ ($\pm 0.04$)   & $1.50$ ($\pm  0.02$)   & $1.69 $ ($\pm 0.05$)  \\
 $2$ &$8$                                  &$Triangular$    & $0.135$   & $2.12$ ($\pm 0.03$)   & $1.23$ ($\pm  0.04$)   & $1.72$ ($\pm  0.08$)  \\
 $$ &$$                                  &$Amorphous$     & $0.146$   & $2.06$ ($\pm 0.02$)  & $1.27$ ($\pm 0.02$)   &$1.62$ ($\pm 0.05$)  \\
\hline $l_x/ \delta l$   & $l_y/ \delta l$       &$Lattice$       & $p_c$     & $\kappa$ &
$\nu_{th}$  & $\nu$ \\ \hline
 $2$ &$2$                                  &$Square$        & $0.405$   & $0.45$ ($\pm 0.07$)   & $1.25$ ($\pm  0.04$)   & $1.22$ ($\pm 0.06$)  \\
 $3$ &$3$                                  &$Square$        & $0.225$   & $0.54$ ($\pm 0.01$)   & $1.31$ ($\pm  0.01$)   & $1.24$ ($\pm 0.05$)  \\
 $5$ &$5$                                  &$Square$        & $0.085$   & $0.54$ ($\pm 0.03$)   & $1.31$ ($\pm  0.02$)   & $1.33$ ($\pm 0.06$)  \\
\end{tabular}
\end{ruledtabular}
\end{table*}
%
%
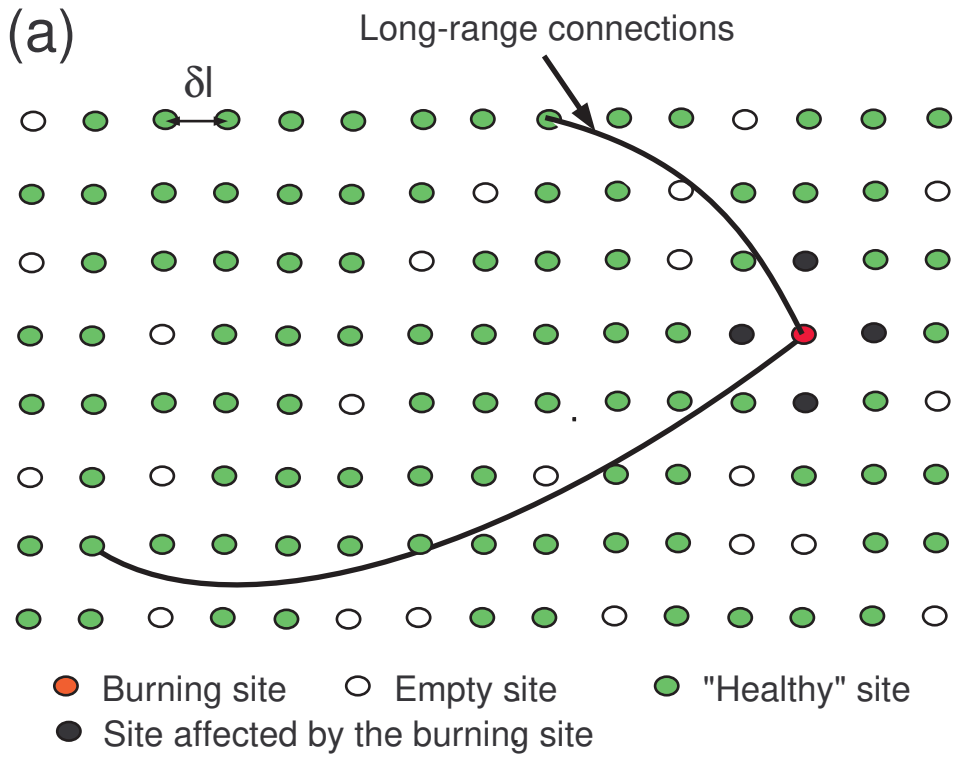
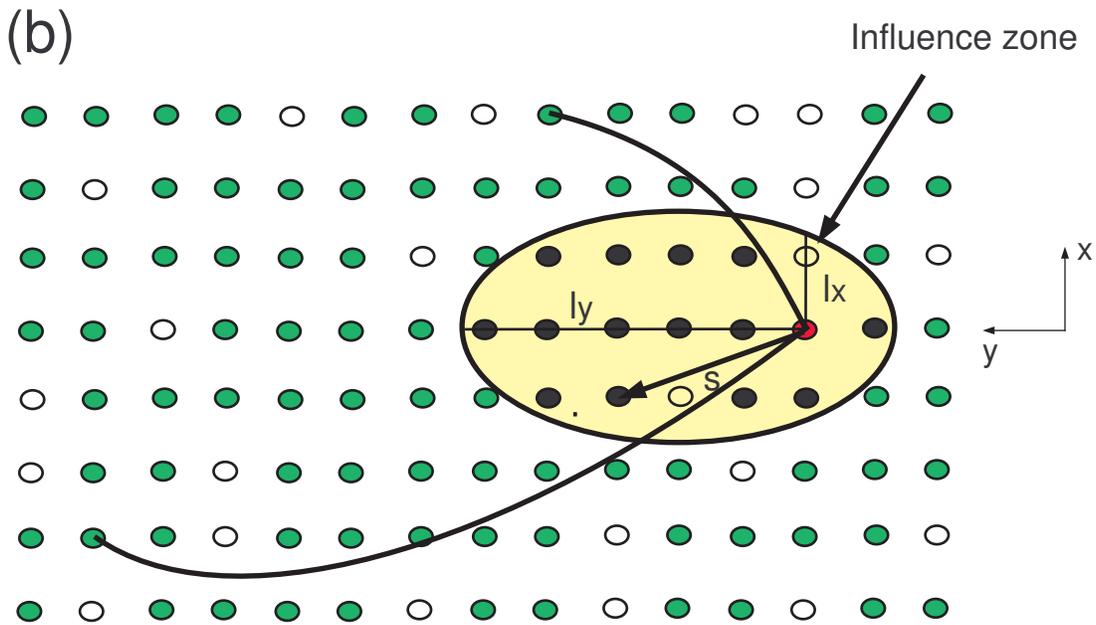
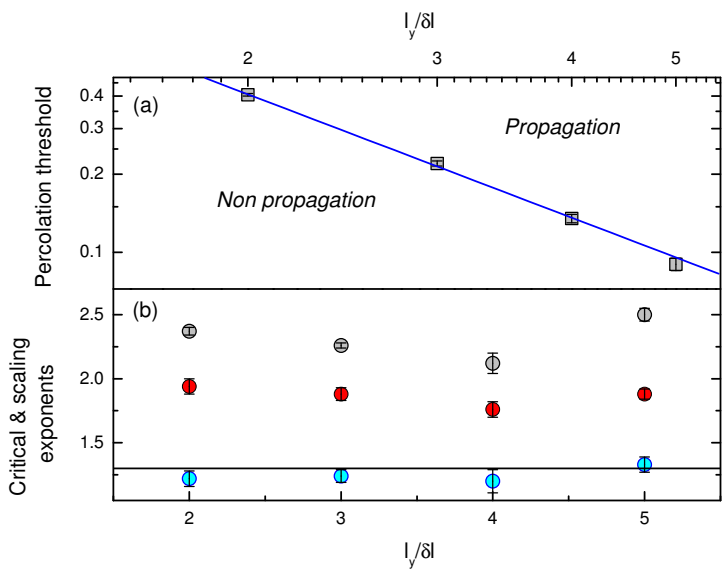
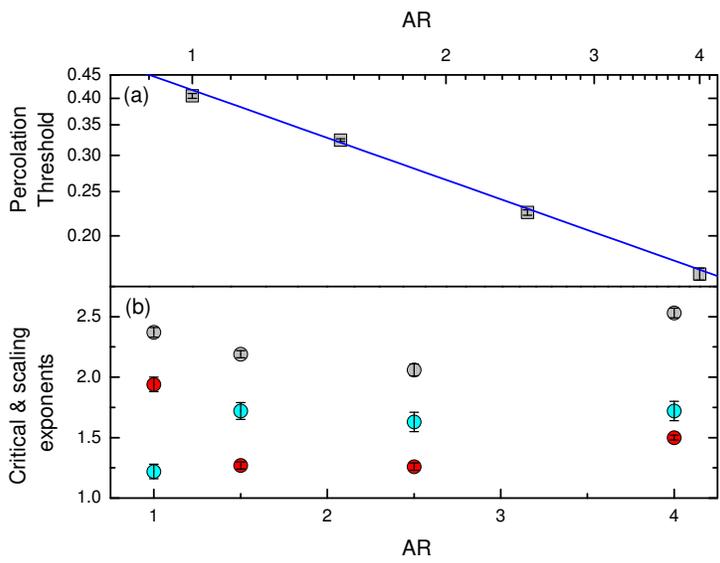
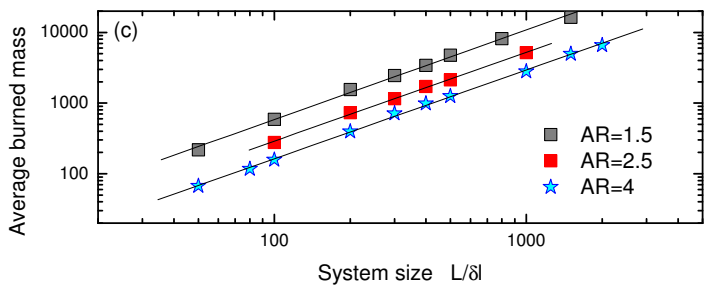
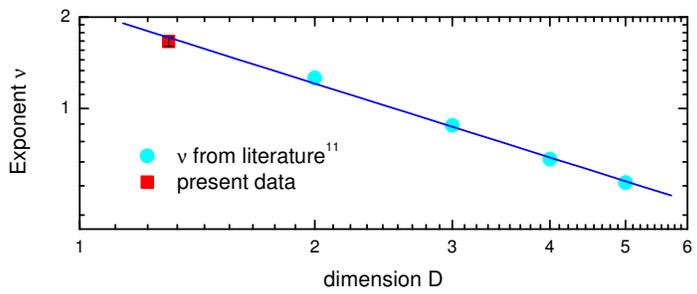
\end{document}